\documentclass[a4paper]{easychair}

\usepackage{doc}
\usepackage{makeidx}
\usepackage{amsmath}
\usepackage{mathtools}
\usepackage{graphicx}
\usepackage{caption}
\usepackage{subcaption}

\newcommand{\easychair}{\textsf{easychair}}

\hypersetup{colorlinks,pdfborder={0 0 0},urlcolor=blue}

\begin{document}

%
\title{The Influence of Decoding Accuracy on Perceived Control: A Simulated BCI Study}


\titlerunning{The {\easychair} Class File}

%
\author{
Pouyan R. Fard\inst{1}
\and
   Moritz Grosse-Wentrup\inst{2}
}

\institute{Graduate School of Neural Information Processing, University of T\"ubingen, T\"ubingen, Germany
  \email{pouyan.rafieifard@student.uni-tuebingen.de}
\and
Max Planck Institute for Intelligent Systems, Dept. Empirical Inference, T\"ubingen, Germany
   \email{moritz.grosse-wentrup@tuebingen.mpg.de}\\
 }

\authorrunning{Mokhov, Sutcliffe, Voronkov and Gough}

\clearpage

\maketitle

\begin{abstract}
Understanding the relationship between the decoding accuracy of a brain-computer interface (BCI) and a subject's subjective feeling of control is important for determining a lower limit on decoding accuracy for a BCI that is to be deployed outside a laboratory environment. We investigated this relationship by systematically varying the level of control in a simulated BCI task. We find that a binary decoding accuracy of 65\% is required for users to report more often than not that they are feeling in control of the system. Decoding accuracies above 75\%, on the other hand, added little in terms of the level of perceived control. We further find that the probability of perceived control does not only depend
on the actual decoding accuracy, but is also in influenced by whether subjects successfully complete the given task in the allotted time frame.
\end{abstract}


%
%

\pagestyle{empty}

\section{Introduction}
\label{sect:introduction}

Brain-computer interfaces (BCIs) are used in a variety of context, ranging from human-computer interaction systems \cite{vandelar2013} over stroke rehabilitation \cite{wentrup2011} to communication for people with severe disabilities \cite{birbaumer1999}. Depending on the context, the performance of a BCI may be assessed in different ways. When using a BCI as a communication device, the \textit{actual} level of  control over the system is often considered more important than the \textit{perceived} level of control. In human-computer interaction systems as well as in the context of stroke rehabilitation, on the other hand, the perceived level of control may be more important to promote interaction with the system than the actual level of control. This raises the question which level of actual and perceived control over a BCI has to be achieved in order to achieve a given task, and how these two measures are related.\\
\indent It is usually assumed that above chance-level decoding accuracy is of little use in BCIs, and that users need to achieve at least 70\% accuracy in a binary decision system in order to reliably to communicate with the system \cite{birbaumer2005}. The level of acceptable and desired accuracy of a BCI based on SSVEPs has been investigated in \cite{ware2010}. Here, it was found that an accuracy level of at least 77\% was desired. In more recent work, the relationship between the actual and the perceived level of control has been investigated in a 2D control navigation game \cite{vandelar2013}. The authors of this work used a traditional keyboard input in an online game and varied the actual level of control. They reported a linear relationship between the actual- and perceived level of control, and a 
non-linear relationship between the perceived level of control and the extent  of user frustration.\\
\indent In this paper, we extend this line of work. We designed a simulated BCI system to study the relationship between perceived- and actual control, i.e. BCI decoding accuracy. By varying the level of control of a binary choice 1-D navigation game with keyboard input, we simulated the decoding and feedback accuracy of a typical BCI system. We then questioned subjects on their subjective feeling of control over the system. Our experimental results indicate that an actual decoding accuracy of at least 65\% is required for users to report more often than not that they were feeling in control of the system. Above 75\%, on the other hand, further increases in terms
of actual control had little influence on the level of perceived control. Furthermore, we found the perceived level of control to strongly depend on whether users successfully completed the given task in the allotted time frame.

\begin{figure}[tb]
	\begin{centering}
	\includegraphics[width=0.7\textwidth]{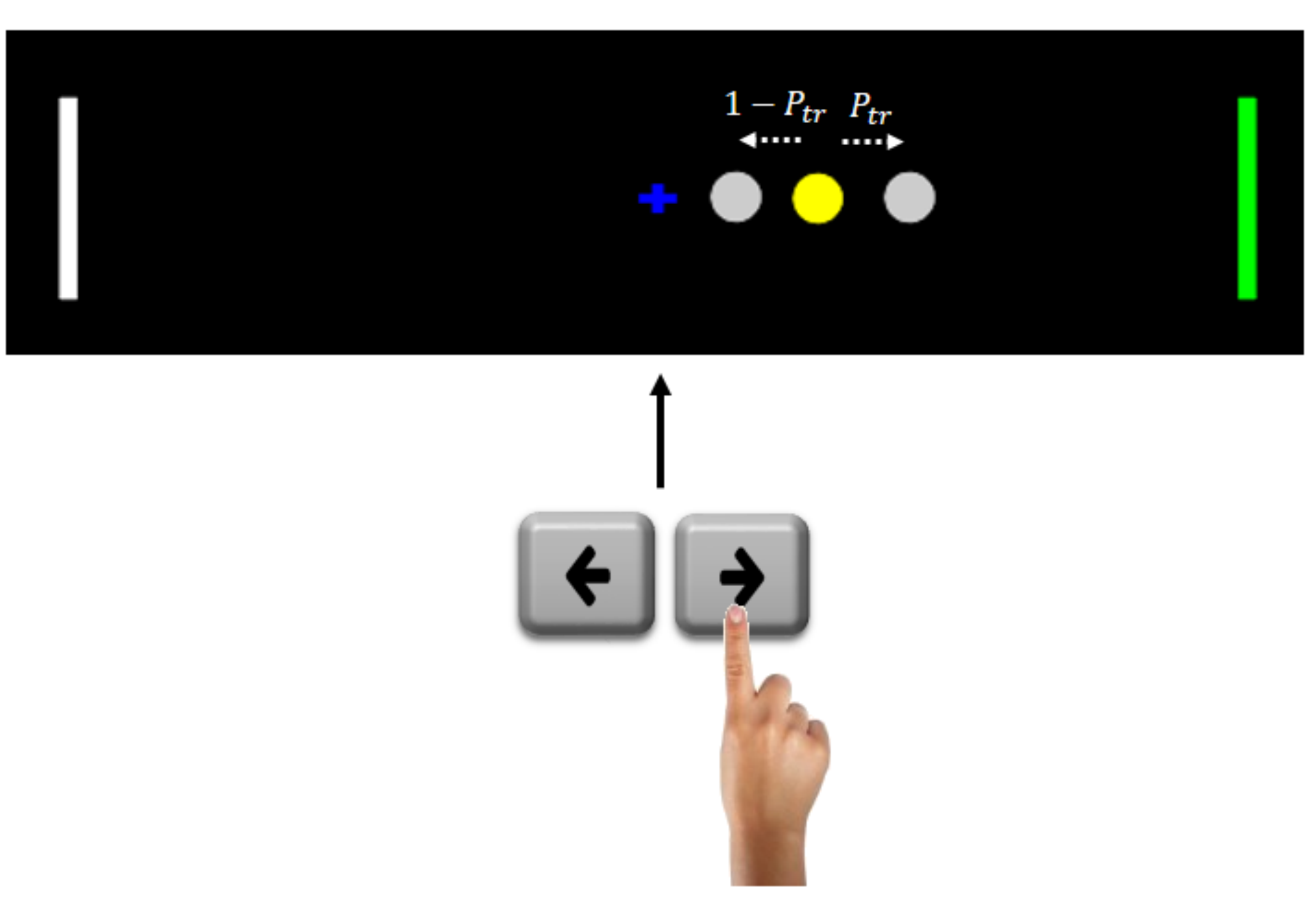}
	\caption{Illustration of one step in the experiment: The condition of the trial is right target. At each time step, the user chooses a direction for the movement (in this case right) and the ball moves to the same direction according to a number drawn from a Bernoulli distribution with success probability of $P_{tr}$ or to the other direction with the probability of $1-P_{tr}$.}
	\label{fig1}
	\end{centering}
\end{figure}

\section{Methods \& Experimental Design}
\label{sect:Methods}

Twenty subjects (nine female) were recruited to participate in a study in which they were asked to navigate a ball to a chosen target using a 1-D binary choice navigation game. All stimuli were shown on a computer monitor located approximately 1.5 meters in front of the subjects. Two grey target bars were shown on the left- and right-hand side of the screen, one of which was selected pseudo-randomly in each trial as the current target by turning it green (Fig.  \ref{fig1}). At the beginning of each trial, a ball was positioned in the center of the screen. The subjects were instructed to move the ball onto the current target by pressing either the ``left''- or the ``right'' key on a computer keyboard in a cued manner. Specifically, the subjects were instructed to press a key within two seconds after the ball turned yellow. Upon a key press the ball turned grey. After three seconds, which simulated a typical delay of a BCI system, the ball moved one step into the desired- or into the opposite direction, depending on a binary number drawn from a Bernoulli distribution with parameter $P_{tr}$ (subsequently termed the actual level of control). This process was repeated until the ball reached the current target (which required a minimum of 11 steps into the correct direction) or if the length of a trial exceeded two minutes. In each trial, the parameter $P_{tr}$ was chosen pseudo-randomly from 11 equally spaced values ranging from 0.5 to 1.\\
\indent Following each trial, subjects were asked to answer the following question: \textit{Do you have the impression that your actions had at least some influence over the ball's movements?} The subjects responded with ``Yes'' or ``No'' by pressing the ``left'' or ``right'' key on the keyboard. Pressing a response key triggered the next trial. One session consisted of three blocks of 11 trials. The pseudo-random selection of the actual level of control for each trial was designed
such that each probability was repeated three times within one session. \\
\indent In order to investigate the relationship of the actual- and the perceived level of control, we computed the percentage of ``Yes''- and ``No''-responses across subjects as a function of the actual level of control. In a second analysis, we split the subjects' responses into two classes depending on whether the current target was or was not reached within the two-minute time limit of a trial.

\begin{figure}[t!]
	\centering
		\begin{subfigure}[t]{0.48\textwidth}
			\includegraphics[width=\textwidth]{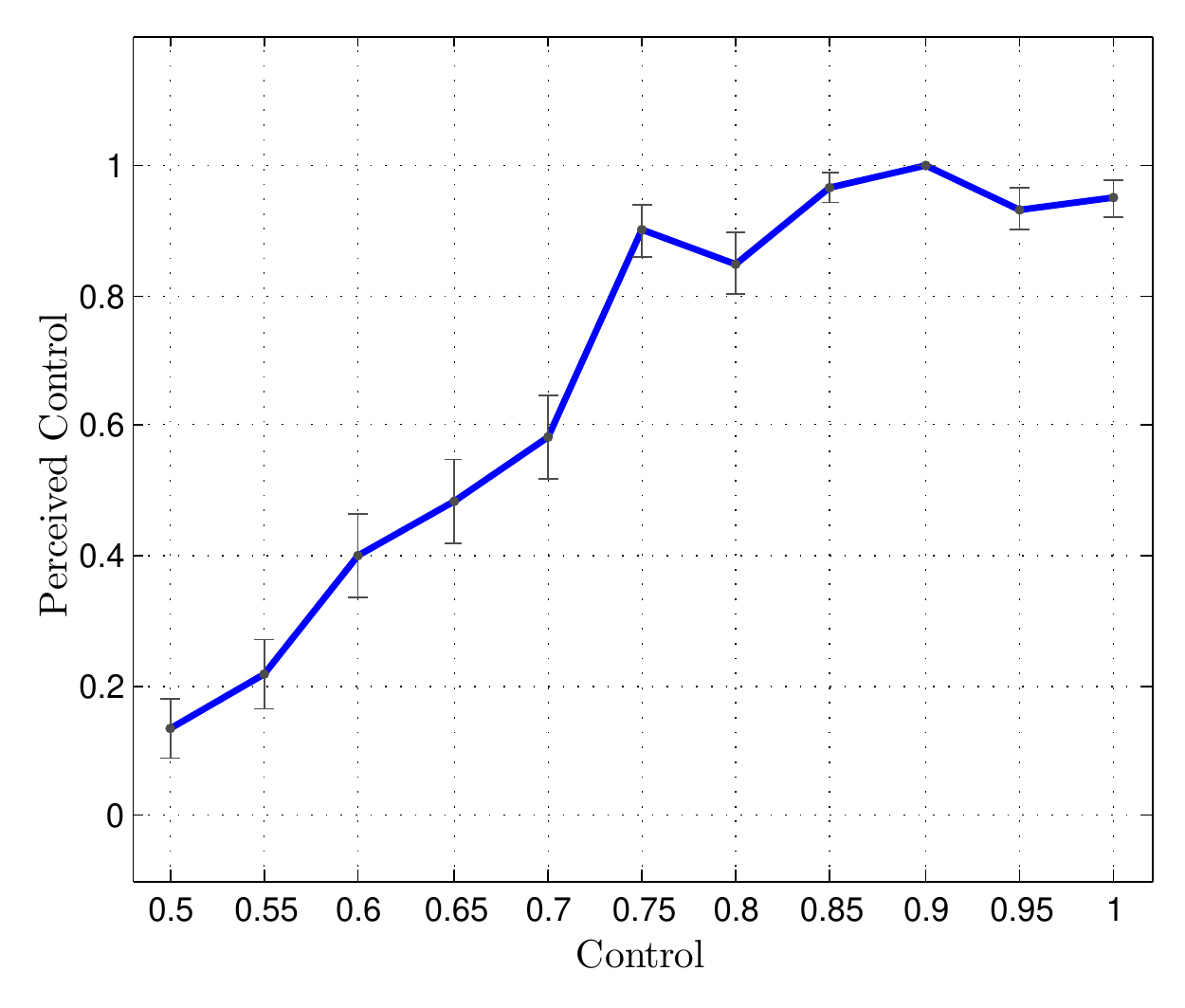}
			\label{fig:easychair-logo}
		\end{subfigure}
		~
		\begin{subfigure}[t]{0.48\textwidth}
			\includegraphics[width=\textwidth]{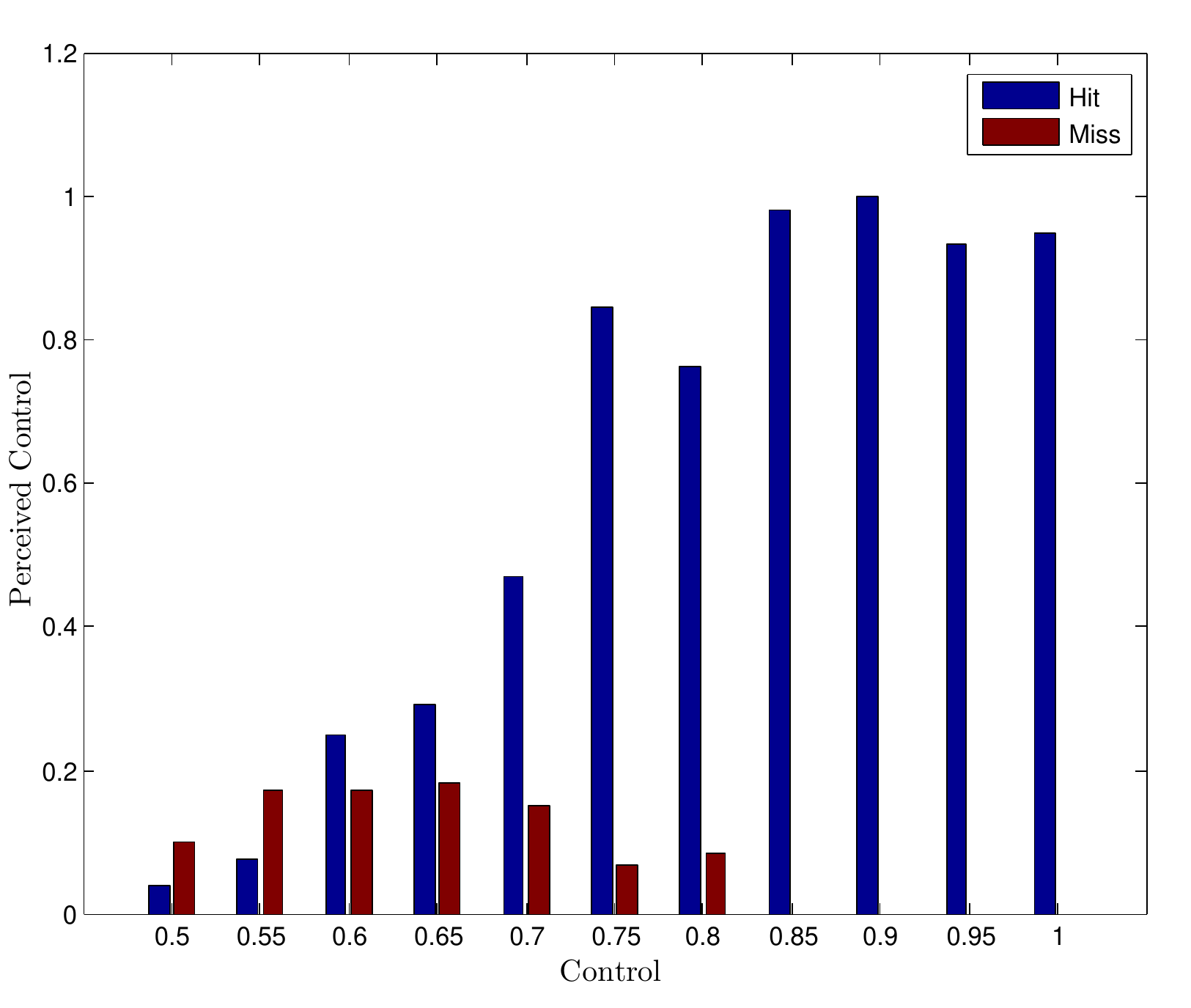}
			\label{fig:easychair-logo}	
		\end{subfigure}
		\caption{A. Perceived control of the users varied with actual levels of control, with error bars indicating the standard error. B. Perceived control of the users split according to the outcome of a trial}\label{fig:animals}
		\label{figure2}
\end{figure}
\section{Experimental Results}
\label{sect:Results}
The probability of a subject feeling in control of the ball is plotted in Fig.  \ref{figure2}.A as a function of the actual level of control. This figure displays a roughly linear relationship between the actual- and the perceived level of control in a range from 50\% (chance-level) to 75\%. Above 65\% of actual control, subjects report more often than not that they perceived some control over the system. Above 75\% of actual control we observe a saturation effect, with higher levels of actual control only resulting in minor improvements in the level of perceived control. Subjects responded in roughly 18\% of trials that they had some feeling of control even when the actual level of control was on chance-level.\\
\indent Fig. \ref{figure2}.B displays the same relationship as Fig. \ref{figure2}.A with trials split up according to whether the target was reached (shown in blue) or the maximum time ran out (shown in red). We found both curves to be fundamentally different. When subjects did not reach the target bar within the allotted time frame, they primarily reported not being able to control the ball. The actual level of control appeared to have little influence on this perception. Even at an actual level of control of 80\% subjects rarely reported to feel in control of the ball. If subjects successfully reached the target, on the other hand, they were far more likely to perceive some level of control over the balls' movements.

\section{Discussion \& Conclusions}
\label{sect:Conclusions}

In this work, we have provided a quantitative analysis of the often used rule of thumb that a decoding accuracy of 70\% is required for a BCI to be deployable outside a laboratory setting. Our results indicate that 65\% of actual decoding accuracy is the lower limit to ensure that subjects more often than not perceive to be exercising some control over the system. Decoding
accuracies above 75\%, on the other hand, added little in terms of perceived control. \\
\indent Our results further indicate that the subjective feeling of being in control is highly influenced by whether subjects successfully complete a given task. In fact, subjects often have the perception of being in control when they successfully complete a trial even for low levels of actual control. Conversely, even for high levels of actual control they rarely reported feeling in control if they did not manage to complete the task in the given amount of time. This observation is in agreement with previously published results on the effect of a positive feedback bias on BCI performance \cite{barbero2010} and emphasizes the importance of providing supportive feedback in order to promote interaction with a BCI system \cite{lotte2013}.\\
\indent It remains an open question, however, which levels of actual- and perceived control are appropriate in a given context. For instance, it has been argued that a high level of perceived control is essential for BCI-based stroke rehabilitation \cite{wentrup2011}. Our results thus indicate that BCI-tasks in stroke rehabilitation should be designed such that patients are likely to complete the task in a given time frame, as this is more likely to result in an accurate estimate of the actual level of control over the system.

\bibliographystyle{ieeetr}
\bibliography{easychair}
\end{document}